# Broadband terahertz time-domain spectroscopy and fast FMCW imaging: principle and applications [*]

Yao-chun Shen(沈耀春)[1][†], Xing-yu Yang(杨星宇)[1] and Zi-jian Zhang(张子健)[1]

[1] *Department of Electrical Engineering and Electronics, University of Liverpool, Liverpool L69 3GJ, UK*

We report a broadband terahertz time-domain spectroscopy (THz-TDS) which enables twenty vibrational modes of adenosine nucleoside to be resolved in a wide frequency range of 1–20 THz. The observed spectroscopic features of adenosine are in good agreement with the published spectra obtained using Fourier Transform Infrared Spectroscopy (FTIR) and Raman spectroscopy. This much extended bandwidth leads to enhanced material characterization capability as it provides spectroscopic information on both intra- and inter-molecular vibrations. In addition, we also report a low-cost Frequency Modulation Continuous Wave (FMCW) imaging system which has a fast measurement speed of 40,000 waveforms per second. Cross-sectional imaging capability through cardboard has also been demonstrated using its excellent penetration capability at a frequency range of 76–81 GHz. We anticipate that the integration of these two complementary imaging technologies would be highly desirable for many real-world applications because it provides both spectroscopic discrimination and penetration capabilities in a single instrument.

**Keywords:** THz-TDS, FMCW radar, adenosine nucleoside

**PACS:** 87.14.gf, 82.80.Gk, 78.47.J-, 84.40.-x

## 1. Introduction

Terahertz (THz) region of the electromagnetic spectrum spans the frequency range between microwave and mid-infrared. It is generally considered to be between 0.1 THz and 10 THz, although sometimes its definition is somewhat arbitrary ranging from 30 GHz to 30 THz overlapping with millimeter-wave (mmWave) and mid-infrared regions. The center portion of the THz region offers a unique combination of many remarkable properties. Firstly, THz radiation gives rise to characteristic spectroscopic 'fingerprints' for many crystalline materials, making THz spectroscopy a useful tool for material characterization. Secondly, THz radiation can penetrate into most polymer and clothing materials, thus THz radiation can be used to image internal structures of a sample. Thirdly, THz radiation is safe to use as its photon energy is millions of times smaller than that of X-rays. These attractive properties and the enormous inherent potential of the THz technology have led to rapid development of THz systems that in turn has opened up many exciting opportunities in academic research and industrial application.[1–4]

[*] The work is partially support by the Royal Society and Natural Science Foundation of China (NSFC) International Exchanges Cost Share (IEC\NSFC\181415)
[†] Corresponding author. E-mail: ycshen@liverpool.ac.uk

Both pulsed and continuous wave (CW) THz technologies have been investigated where pulsed measurements yield more information whilst CW imaging allows faster measurements.[4] The core technology behind the pulsed THz time-domain system is the coherent generation and detection of short pulses of broadband THz radiation by using an ultrafast femtosecond laser. A number of techniques including ultrafast switching of the photoconductive antenna,[5–7] bulk electro-optic rectification using non-linear crystal materials,[8, 9] and laser-induced breakdown from air[10] and water film,[11] have been explored for generating short pulses of THz radiation. On the coherent detection of THz radiation, both photoconductive receiver antenna[12] and non-linear crystal such as ZnTe (via electro-optic sampling)[13] has been used in a pump-and-probe fashion where the optical probe beam is from the same femtosecond laser. This coherent generation and detection scheme allows the transient electric field, rather than the intensity of the THz radiation, to be measured directly. This not only yields THz spectrum with far better signal to noise ratio and dynamic range as compared with the Fourier Transform Infrared Spectroscopy (FTIR) method, but also preserves the time-gated phase information, upon which THz ranging and imaging has been developed for characterizing the internal structures of a sample quantitatively and non–destructively. To date, both THz time-domain spectroscopy and imaging systems are commercially available, but most THz systems have a limited usable spectral range of about 3 THz. Broader spectral coverage is highly desirable as it would not only provide more spectroscopic signatures for better material characterisation but also enable quantitative imaging of thinner layers with higher spatial resolution. Here we report a pulsed THz time-domain system with an extended usable spectral range up to 20 THz.

The principle of continuous wave (CW) THz imaging has existed for several decades.[14] In the past two decades, there has been a growing interest in developing CW THz and mmWave imaging technology primarily for security scanning, poor-weather navigation and military applications.[15] Unlike pulsed THz imaging where a femtosecond laser and complex optics are necessary, CW imaging system can be made purely electronic for both generation and detection of THz radiations. Therefore, CW imaging system is usually more compact, simpler, and the measurement is faster, although conventional CW imaging only yields intensity data and it does not provide any depth or frequency-domain information.[16] More recently, with its emerging application in assisted and autonomous driving, there has been enormous interest in the development and applications of the Frequency Modulated CW (FMCW) radar technology.[17] The FMCW radar allows the range and velocity information of moving objects to be detected simultaneously in real time, thus providing crucial information for the control system of the self-driving vehicle to enable safe and collision-free cruise control.[18] In this work, we report the development of a FMCW imaging system using the low-cost off-the-shelf components developed for automobile industry.

## 2. Broadband THz time-domain spectroscopy

Figure 1 shows the experimental arrangement for coherent generation and detection of broadband THz radiation schematically.[19–21] In brief, the output of a femtosecond laser (12-fs pulse width, 800-nm centre wavelength, 76-MHz repetition rate, 300-mW average power) is split using a 90:10 beam-splitter into two parts: 1) a 270 mW pump beam is focused onto the surface of a biased photoconductive antenna for THz generation; 2) a 30 mW probe beam is focused onto a 20-μm-thick ZnTe crystal glued on the top of a 1-mm-thick wedged ZnTe crystal for electro-optic detection.[8] The photoconductive emitter comprised two vacuum-evaporated NiCr/Au electrodes separated by a gap of 0.4 mm, deposited on a 1-μm-thick low-temperature (LT) GaAs layer grown on a 0.53-mm-thick undoped GaAs substrate.[12] A bias voltage of ±120 V, modulated at 31 kHz, was applied across the emitter. Under the applied electric field, the photo-generated carriers in GaAs crystal by the femtosecond laser pulse will produce a transient current which in turn generates a pulse of electromagnetic radiation at THz frequencies. The generated THz radiation is collected and focused into a sample by using a pair of off-axis parabolic mirrors. After transmitting through the sample, the THz radiation is collimated and focused, using another pair of off-axis parabolic mirrors, onto the ZnTe crystal for THz detection via electro-optic sampling using the probe beam from the same ultrafast laser. In this way, the time-resolved electric field of the THz pulse could be recorded by scanning the time-delay between the THz pulse and the near-infrared probe beam using a variable delay stage.

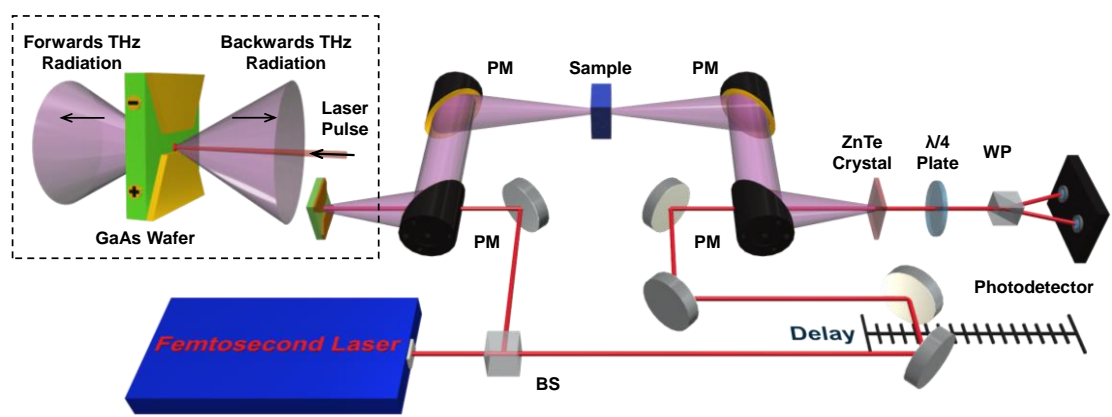

**Fig. 1.** Experimental arrangement for coherent THz generation and detection. Insert: Generated THz radiation is emitted from both sides of the GaAs wafer and the backwards THz radiation is collected in our measurements. (BS – beam-splitter; PM – off-axis parabolic mirror; WP – Wollaston prism)

In contrast to conventional experiments where the THz radiation was collected forwards (that is after being transmitted through the semiconductor substrate), we collected the THz radiation backwards (on the same side of the incident pump laser beam, see the insert of Fig. 1). As a result, the absorption and dispersion of the THz pulses in the GaAs substrate were minimized.[19, 20] Fig. 2(a) and 2(b) show typical resulting temporal THz waveforms that were measured using a 200-μm-thick GaP and a 20-μm-thick ZnTe crystal, with Fig. 2(c) and 2(d) showing their corresponding FFT amplitude spectra in logarithm and linear scales, respectively. In time-domain (Fig. 2(a) and 2(b)), the 200-μm-thick GaP detector provides about eight times larger signal than the 20-μm-thick ZnTe detector. In frequency-domain (Fig. 2 (c) and 2(d)), the 200-μm-thick GaP detector still provides larger signal than the 20-μm-thick ZnTe detector in the lower frequency range (up to about 6 THz). However, in the higher frequency range of 6–20 THz, the 20-μm-thick ZnTe detector provides larger signal. The signal-to-noise ratio is greater than 30 at 10 THz and it is still above 3 for frequency up to 18 THz, thus enabling reliable spectroscopic measurement in this newly extended frequency range.

As a demonstration, the THz transmission spectrum of polycrystalline adenosine was measured using the 20-μm-thick ZnTe detector (shown in Fig. 3). Adenosine is one of four nucleosides found in RNA and DNA, and polycrystalline samples contain an extensive network of inter-molecular hydrogen-bonds.[22] Samples for THz measurements were prepared by forming a thin layer of finely milled adenosine powder between two transparent cling films. As shown in Fig. 3, our THz measurements provide a much-extended spectral coverage of 1–20 THz and reveal twenty vibrational modes. Note some THz-TDS systems may contain measurement artefacts, for example, standing wave may lead to periodic spectral ripple artefacts. However, as shown in Fig. 2(d), the spectrum measured in our THz-TDS system shows no spectral ripples or features when there is no sample presented. Also, the observed spectral features in Fig. 3 are not of periodic nature either, thus these spectral features are truly from adenosine sample. In addition, most THz spectroscopic features of many materials of biological significance are relatively weak, particularly when measured at room temperature such as the ones reported here. Nevertheless, about twenty spectral features can still be identified from Fig. 3. Table 1 summarises the spectral features observed in this work, together with published spectral features of adenosine obtained using FTIR and Raman spectroscopy as well as neutral inelastic scattering (NIS) spectroscopy methods.[23-25] In general, there is good agreement between THz measurements and other well-established methods, a further validation that these observed spectral features are characteristic fingerprints of adenosine sample.

Note that most commercial THz-TDS product has a typical bandwidth of about 3–4 THz. This, together with its high price tag, has been a major limiting factor for the wide uptake of THz-TDS technology particularly in industry sectors where the performance-price ratio is of major impact. There has been considerable interest and research work to extend the bandwidth of THz-TDS. One method is the optical excitation of THz radiation using new type of nonlinear crystal materials.[8, 26] The other method is to improve the performance of photoconductive antenna by using novel antenna structure and new substrate materials. GaAs semiconductor crystal provides good photo-to-THz conversion efficiency and absorbs little THz radiation below 3 THz thus it is the material of choice for most photoconductive antenna used in commercial THz-TDS systems. However, GaAs crystal has a resonance absorption peak around 8 THz (Fig. 2 (c) and 2(d)), leading to strong absorption of THz radiation above 4 THz – the closer to 8 THz, the stronger the absorption. Therefore, the bandwidth of most commercial THz-TDS products is limited to be about 3–4 THz.

In this work, a backward collection configuration (Fig. 1) was used and this allows the THz radiation generated from the photoconductive antenna to be collected and transmitted to the THz receiver without passing through the GaAs substrate. In this way, the THz absorption by the GaAs substrate is completely avoided, thereby providing an extended bandwidth up to 20 THz. The basic principle was reported previously by the same author[19] and more recently was used by research teams in Japan and France who have achieved a 20 THz broadband generation using GaAs interdigitated photoconductive antennas.[20] In 2020 a research team in Germany reported a bandwidth of 70 THz by using the implanted Ge crystal (which has no resonance absorption in the THz frequency range of interest), rather than GaAs crystal (which has strong resonant absorption in the THz frequency range of interest), as the substrate material of photoconductive antenna.[21] This is a further demonstration that the bandwidth of a photoconductor THz antenna is ultimately limited only by the duration of the excitation and detection laser pulses if one could minimize the THz absorption by the antenna substrate. On the other hand, GaAs photoconductive THz antenna is advantageous over intrinsic Ge photoconductive antenna because GaAs has higher carrier mobility and shorter carrier lifetime thus better photo-THz conversion efficiency. In this work, using GaAs photoconductive antenna in a backward collection scheme, we achieved an effective spectroscopic measurement range of 20 THz, far exceeding the typical 3–4 THz measurement range of conventional commercial THz-TDS products. Our approach does not add any additional cost to the THz-TDS system thus it has higher performance-to-price ratio. We hope that our work will attract more researchers to develop further this technology, making the ultra-broadband THz-TDS more widely and readily available.

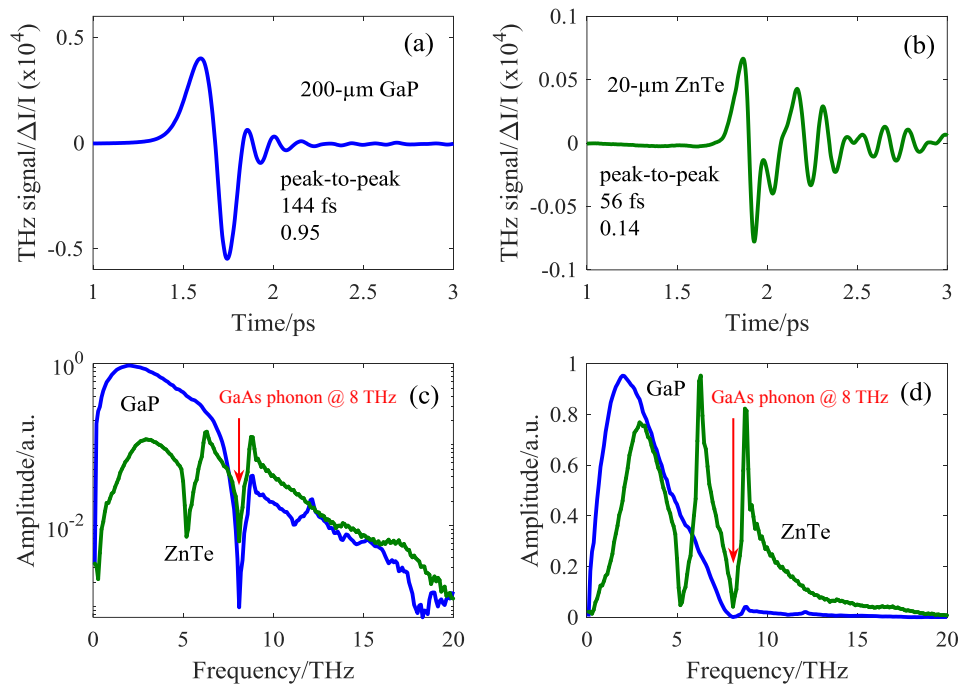

**Fig. 2.** Temporal THz signal measured using a 200-µm-thick GaP (a) and a 20-µm-thick ZnTe crystal (b), and their corresponding FFT amplitude spectra in logarithm (c) and linear scales (d). The spectral dips at 5.3 THz and 8 THz are caused by ZnTe detector (phonon mode @ 5.3 THz) and GaAs crystal (phonon mode @ 8 THz), respectively.

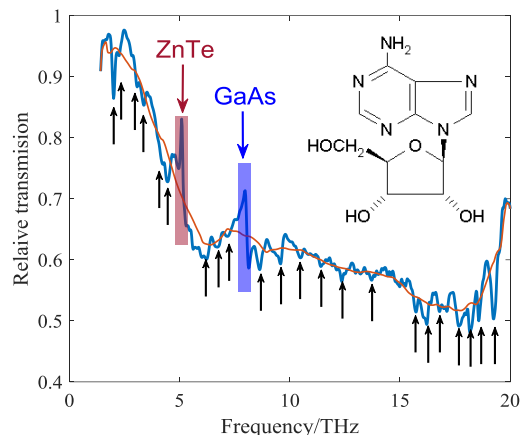

**Fig. 3.** Broadband transmission spectrum of an adenosine sample. Two shaded areas centred at 5.3 and 8 THz contain singularities caused by the TO phonons of the ZnTe and GaAs crystals, respectively. Arrows indicate vibrational modes at twenty frequencies of 2.0, 2.2, 3.0, 3.4, 4.2, 4.5, 6.2, 6.7, 7.2, 8.6, 9.6, 10.5, 11.4, 12.4, 15.7, 16.2, 16.8, 17.7, 18.2, 19.2 THz, with a frequency resolution of 0.075 THz. The insert shows the molecular structure of adenosine ($C_{10}H_{13}N_5O_4$). The red curve is a moving average of the transmission spectrum, as a guide to the eye.

**Table 1.** Spectral signatures of Adenosine (THz).

| This work | Published Work (in Refs. [23 –25]) | | | |
|---|---|---|---|---|
| THz-TDS | FTIR | FT-Raman | Raman | NIS |
| 19.2 | 19.2 | 19.2 | 19.2 | 19.4 |
| 18.2 | | | | |
| 17.7 | 17.8 | 17.6 | 17.6 | 17.5 |
| | | 17.2 | 17.1 | |
| 16.8 | 16.5 | 16.6 | 16.5 | |
| 16.2 | 16.2 | 16.1 | 16.1 | |
| 15.7 | 15.8 | 15.8 | 15.8 | 15.8 |
| 12.4 | 12.4 | 12.4 | 12.4 | 12.0 |
| 11.4 | 11.7 | 11.5 | 11.5 | |
| 10.5 | 10.5 | 10.5 | 10.5 | |
| 9.6 | 9.6 | 9.6 | 9.6 | 9.4 |
| 8.6 | 8.6 | 8.6 | 8.6 | 8.3 |
| 7.2 | | 7.2 | 8.1 | |
| 6.7 | 6.8 | 6.8 | | 6.6 |
| 6.2 | | 6.2 | | |
| 4.4 | 4.4 | | | 4.4 |
| 4.2 | | 4.2 | | |
| 3.4 | | 3.4 | 3.4 | |
| 3.0 | | 3.1 | 3.1 | |
| 2.2 | | 2.4 | 2.4 | |
| 2.0 | 2.0 | 2.1 | 2.1 | |

Spectral features in the mid-infrared region are dominated by intra-molecular vibrations of sample molecules whilst spectral features in THz region are dominated by inter-molecular vibrations, corresponding to motions associated with coherent movements of large numbers of atoms and molecules. Such collective phonon modes only exist in materials with periodic structure. The THz time-domain spectroscopy reported here greatly extended the upper end of the THz spectrum from about 3–4 THz to 20 THz, providing access to both intra- and inter-molecular vibrational modes. On the other hand, this THz time-domain spectroscopy system requires an expensive femtosecond laser, complex optics and opto-mechanics such as time-delay scanner, and the measurement time is long (about one spectrum per minute) in order to have a reasonable signal to noise ratio over the whole spectral range of 1–20 THz. For applications that fast measurement speed is essential, the FMCW imaging technology introduced in the next section would be a more suitable choice.

## 3. FMCW imaging

As shown in Fig. 4, for pulsed THz time-domain imaging, a pulse of THz radiation is sent to a sample under investigation. The THz radiation reflected from the sample is then measured as a function of time. The surface reflection peak and the interface reflection peaks/valleys are separated in time in the measured THz waveform (Fig. 4(a)). The principle of a FMCW radar system is similar, albeit the measurement is implemented in frequency domain.[17] As shown in Fig. 4(c), a transmit antenna of the FMCW radar sends a continuous wave with a time-varying frequency $f(t)$. Assuming sawtooth modulation was used, e.g., the transmitted frequency increases linearly as a function of time during sweep period, we have

$$f(t) = f_0 + \alpha t \tag{1}$$

where $f_0$ is the starting frequency, $\alpha$ is the chirp rate which is defined as $B/T$, $B$ is the bandwidth and $T$ is sweep period (Fig. 4(d)). The transmitted signal in the $i$-th sweep period can be written as:

$$E_{TX}(t) = E_{T0} \cos(\phi_0 + 2\pi(f_0(\tau + (i-1)T) + \frac{\alpha \tau^2}{2})) \text{ where } (0 \leq \tau \leq T) \tag{2}$$

The transmitted wave, after being reflected by an object, is received by a receiver antenna. Assuming the object is at an initial distance of $R$ away from the transmit antenna and is moving away with a relative velocity of $v$. The returned signal from the object will have the same form as the transmitted signal, but with some time delay $t_d$ which can be defined as:

$$t_d = \frac{2(R + v((i-1)T + \tau))}{c} \tag{3}$$

Considering this time delay, the return signal can be expressed as:

$$E_{RX}(t) = E_{R0} \cos(\phi_0 + 2\pi(f_0(\tau - t_d + (i-1)T) + \frac{\alpha(\tau - t_d)^2}{2})) \text{ where } (0 \leq \tau \leq T) \tag{4}$$

According to the FMCW radar principle, the returned signal is mixed with the transmitted signal. By filtering out the sum signal, we have the final phase difference signal:

$$E_m(t) = \frac{E_{T0}E_{R0}}{2}\cos(2\pi f_0 t_d + \frac{\alpha}{2}(2\tau t_d - t_d^2)) \quad \text{where } (0 \leq \tau \leq T) \quad (5)$$

By processing the above signal, one can obtain both the distance and velocity of the object. The distance of the target can be easily found by applying the FFT algorithm over one sweep signal period:

$$R = \frac{f_b c}{2\alpha} \quad (6)$$

where $c$ is the speed of light and $f_b$ is the beating frequency obtained from the FFT amplitude spectrum. For THz imaging of a static sample with a single layer of a thickness $d$, Eq. 3 can be simplified as:[27]

$$t_{d1} = \frac{2R}{c}$$
$$t_{d2} = \frac{2(R+nd)}{c} \quad (7)$$

where $n$ is the refractive index and $d$ is the layer thickness. The range resolution, e.g., the thinnest layer that can be resolved, is determined to be 3.75 cm in air ($n = 1$) by using the following equation:

$$R_{resol} = \frac{c}{2nB} \quad (8)$$

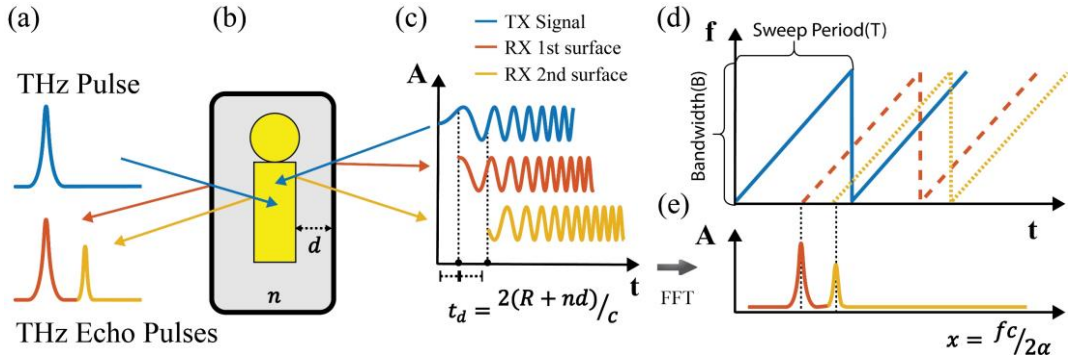

**Fig. 4.** Comparison of the working principles between pulsed THz time-domain imaging and FMCW imaging. (a) Incident THz pulse and returned THz echo pulses where the first peak corresponds to the sample surface reflection and the second peak is the interface reflection from within the sample. (b) Schematic diagram of a sample with a single layer with a thickness of $d$, and a refractive index of $n$. (c) FMCW radar transmitted (TX) and returned (RX) signals. Note the different time delay (phase difference) corresponding to surface and interface reflections. (d) TX and RX frequency-varying signals with sawtooth frequency modulation. (e) The depth profile calculated from the FFT amplitude spectrum of the signal.

Figure 5 shows the schematic diagram of an FMCW imaging system. The core component is a single-chip FMCW radar developed primarily for automobile applications. It integrated all required key components, including self-oscillation circuit (OSC) and mixer, pre-amplifier (PA) for the transmit antenna, low-noise amplifier (LNA) for the receive antenna, the low-pass filter (LPF), the digital signal processing unit (DSU) and the analogue-digital-converter (ADC) into a single-chip. This greatly reduced the complexity and the cost of an FMCW imaging system. A single PTFE lens was used to collect and focus the mmWave radiation onto an imaging target. The mmWave signal returned back from the sample was collected and focused back to the receive antenna using the same lens. In this experiment, objects are placed inside a cardboard box to demonstrate the see-through imaging capability of the system for potential security screening applications. The central wavelength of the mmWave used is 3.8 mm and the spot size of the focused mmWave beam at the sample position is estimated to be 6 mm. The sample is fixed on a motorised translation stage. The cross-sectional image of the target can be obtained by laterally translating the target across the focused radiation. An in-house software (with MATLAB 2019a) is used for image acquisition and image reconstruction.

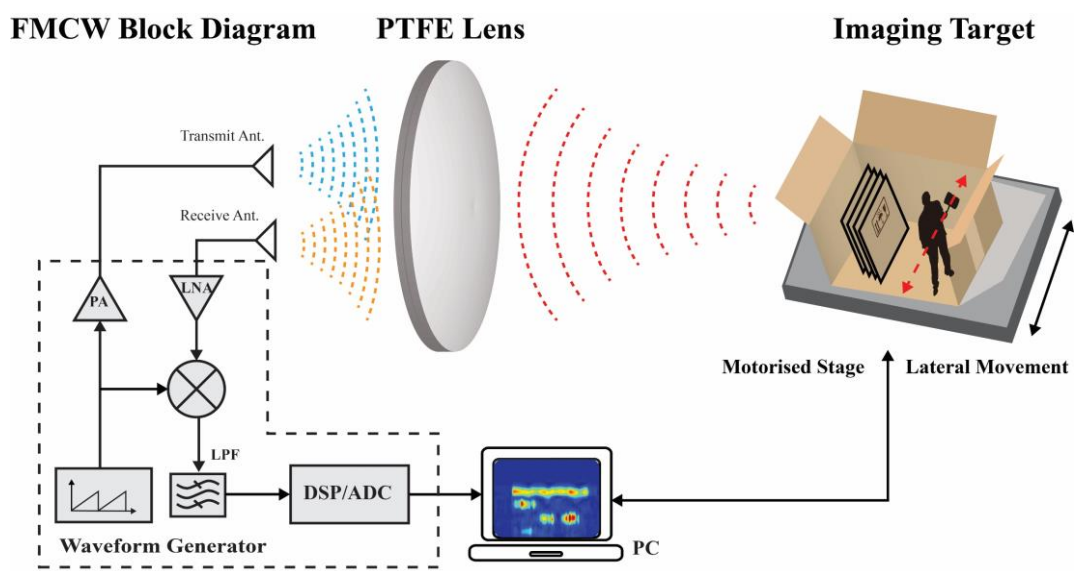

**Fig. 5.** The proof-of-principle FMCW imaging system based on a FMCW single-chip.

As an example, Figure 6 shows cross-sectional images and representative amplitude waveforms of three test samples measured using the FMCW imaging system. These three test samples are: 1) an empty cardboard box; 2) a metal pot placed behind a cardboard; 3) three metal pieces placed at different depths behind a cardboard. In all cases, the objects behind the cardboard can be easily resolved. The achieved image resolution is approximately 6 mm laterally and 38 mm axially. It should be noted that the axial resolution is determined by the limited sweep bandwidth of 5 GHz of the specific FMCW radar chip used and can be greatly improved by increasing the sweep bandwidth of a FMCW system. The ADC in the radar chip has a sampling rate of 40 MHz and the measured signal has 1,000 points in each sweep period, thus enabling an acquisition rate of 40,000 waveforms per second. This is about 1,000 times faster than most pulsed THz time-domain imaging system.

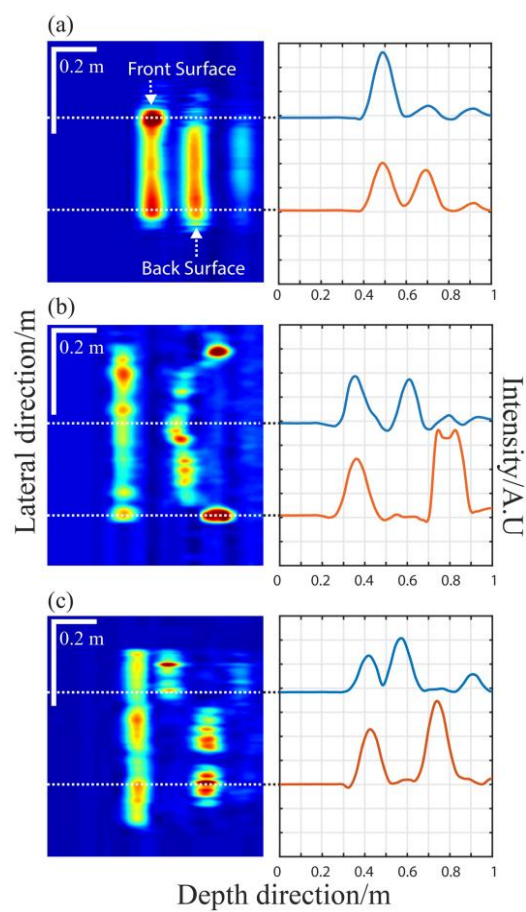

**Fig. 6.** Cross-sectional images and A-scans acquired using the system. (a) an empty cardboard box showing its front and back, (b) a metal pot with a curved surface behind a piece of cardboard, (c) and multiple metal pieces at different depth behind a piece of cardboard.

In order to demonstrate further the penetration capability of the FMCW radar reported here, Figure 7 shows the measured signals after passing through 1, 2, 3, and 4 pieces of cardboards, respectively. Ten repetitions of the same test were carried out and Figure 7(b) shows the signal strength against the number of cardboards, together with the standard deviation of the ten repeated measurements. We note that the measured signal attenuation is not linear to the number of cardboards. This was expected as there will be additional multi-reflection losses at the surface of each cardboard, apart from the absorption losses. Nevertheless, it is clear that an object behind four pieces of cardboards with a total thickness of over 20 mm can still be detected in a single measurement without the need of any averaging. Note that one could improve the signal to noise ratio further by averaging over multiple sweep periods.

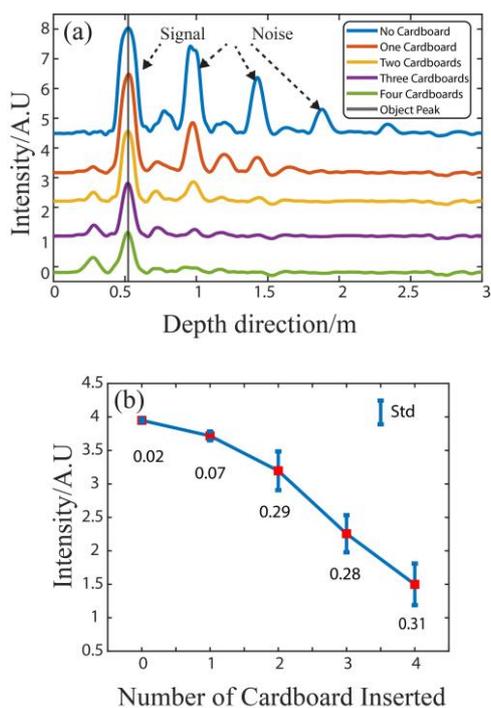

**Fig. 7.** Attenuation analysis. (a) The signal strength of the metal surface attenuation plot. (b) Standard deviation error bar of 10 sets of repeated experiments.

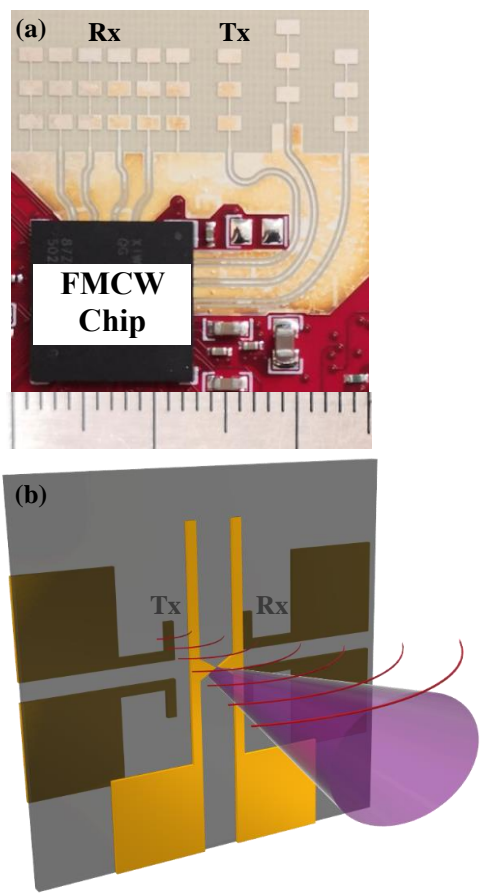

**Fig. 8.** (a) Photograph of the FMCW radar chip together with its transmitter antenna (Tx) and receiver antenna (Rx). (b) Schematic diagram of an integrated THz and FMCW antennas. Note that THz antenna and FMCW antennas are on the different side of the GaAs substrate, thus minimizing possible interference between THz and FXCW measurements.

## 4. Conclusions and future prospects

We reported a pulsed THz time-domain measurement technique capable of obtaining THz spectra in a frequency range of 1–20 THz. Twenty spectral features of adenosine were resolved using such a pulsed THz time-domain spectroscopy. We noted that FTIR could also be used to carry out spectroscopic measurements at the upper end of the THz range. However, FTIR requires detector at cryogenic temperatures whilst THz spectroscopy measurement can be done at room temperature. Furthermore, pulsed THz time-domain measurement method can be applied to measure heated sample because of its time-gated coherent generation and detection scheme whilst FTIR may not be suitable to study heated sample. It is because samples at high temperature will radiate substantial infrared radiation that will lead to large background noise to the measured signal or even saturate the FTIR detector. In addition, Raman spectroscopy method could also be used to obtained spectroscopic information at THz frequencies. However, the optical absorption spectra (THz and infrared) and Raman scattering spectra are in general complementary. Depending on the nature of the vibration, which is determined by the symmetry of the molecule, vibrations may be active or forbidden in the infrared or Raman spectra. Therefore, the ultra-broadband THz spectroscopy reported here provides a unique capability to measure THz spectra at room temperature even for heated samples and this will open up a number of important applications.

In this work, we also reported the development of FMCW imaging technique using the commercially available off-the-shelf components developed primarily for the mass automobile market. The research on microwave and mmWave imaging has a long history, and the basic scientific principles have not changed. However, there have been significant development and advances in engineering innovation and technology integration in recent years. From the perspective of practical applications, the performance-price ratio is an important factor. For example, the mmWave imaging technology developed by Zhang's group has been successfully applied for the inspection of space shuttle, which represents a major milestone of the practical applications of mmWave imaging technology.[28] However, it is prohibitively expensive for many cost-sensitive practical applications. Our FMCW imaging system uses components developed and mass produced for automobile market. The developed system is compact and cost less than $500, excluding a control computer and motorised stage. Therefore, it can be easily integrated into THz-TDS or THz time-domain imaging systems, providing a perfect combination of two technologies with complementary capabilities, e.g., penetration capability and spectroscopic discrimination capability. In addition, there are currently a number of research institutions and companies that are actively promoting the extension of the frequency modulation range from mmWave to THz range. As shown in Fig. 8(a), the core component (FMCW radar chip) is a self-contained integrated circuit (system on chip) with a chip size of about 12mm × 12 mm and the 3 transmitter and 4 receiver mmWave antennas are all fabricated on a printed circuit board (PCB). Furthermore, in our THz-TDS system the THz antenna is on the frontside of GaAs photoconductive antenna (inset of Fig. 1) to minimize absorption at higher THz frequencies. This frees the backside of the GaAs photoconductive antenna where FMCW transmitter and receiver antennas can be integrated (Fig. 8(b)). The

separation of the THz and FMCW antennas on different side of the GaAs substrate is important and advantageous as it minimizes the possible electromagnetic interference between THz and FMCW measurements. Note that at the increased operating frequency, the FMCW transmitter and receiver antennas will have smaller size, making it technically and economically viable to be integrated into the THz photoconductive antenna which is usually small in size to minimize the cost associated with the GaAs crystals. In addition, GaAs substrate used in THz photoconductive antenna is lossless at FMCW operating frequencies and it will provide additional benefits as compared with antennas using PCB substrates that will have increased losses at higher frequencies.

In addition, both mmWave and THz waves are intrinsically harmless to human and both are able to penetrate clothing materials. As demonstrated here, mmWave imaging systems are able to image through clothing materials in real time, whilst broadband THz technology provides unique spectroscopic information for substance discrimination. It is therefore feasible to develop a THz hybrid imaging and sensing system for security screening application by combining the real-time imaging capability of mmWave technology and the spectroscopic discrimination capability of broadband THz technology.